\documentclass[twocolumn,showpacs,aps,groupedaddress,prd,eqsecnum]{revtex4}
\usepackage{amsfonts,latexsym,eucal}
\usepackage[dvips]{graphicx}

\def\beq{\begin{equation}}
\def\eeq{\end{equation}}
\def\bea{\begin{eqnarray}}
\def\eea{\end{eqnarray}}

\def\pa{\partial}
\def\ra{\rightarrow}

\def\tt{\tilde{t}}
\def\tr{\tilde{r}}
\def\tp{\tilde{\phi}}
\def\tpp{{\tp_{~}}'}
\def\tpps{{\tp_{~}}^{\prime2}}
\def\tv{\tilde{V}}

\def\to{\tilde{\omega}}
\def\bp{\mbox{\boldmath$\phi$}}
\def\tE{\tilde{E}}
\def\tQ{\tilde{Q}}

\usepackage{epsfig}

\begin{document}

\thispagestyle{empty}

\title{Gravitating Q-balls in the Affleck-Dine mechanism}
\author{Takashi Tamaki}
\email{tamaki@ge.ce.nihon-u.ac.jp}
\affiliation{Department of Physics, General Education, College of Engineering, 
Nihon University, Tokusada, Tamura, Koriyama, Fukushima 963-8642, Japan}
\author{Nobuyuki Sakai}
\email{nsakai@e.yamagata-u.ac.jp}
\affiliation{Department of Education, Yamagata University, Yamagata 990-8560, Japan}
\date{\today}

\begin{abstract}
We investigate how gravity affects ``Q-balls" with the Affleck-Dine potential 
$V_{AD}(\phi):=\frac{m^2}{2}\phi^2\left[ 1+K\ln \left(\frac{\phi}{M}\right)^2\right]$.
Contrary to the flat case, in which equilibrium solutions exist only if $K<0$, we find three types of gravitating solutions as follows.
In the case that $K<0$, ordinary Q-ball solutions exist; there is an upper bound of the charge due to gravity.
In the case that $K=0$, equilibrium solutions called (mini-)boson stars appear due to gravity; there is an upper bound of the charge, too.
In the case that $K>0$, equilibrium solutions appear, too. In this case, 
these solutions are not asymptotically flat but surrounded by Q-matter. 
These solutions might be important in considering a dark matter scenario in the Affleck-Dine mechanism. 

\end{abstract}

\pacs{04.40.-b, 05.45.Yv, 95.35.+d}
\maketitle

\section{Introduction}
Q-balls \cite{Col85}, a kind of nontopological solitons \cite{LP92}, appear in a large family of field
theories with global U(1) (or more) symmetry. In particular, it has been argued that 
Q-balls with the Affleck-Dine (AD) potential could play important roles in cosmology \cite{AD}.
For example, Q-balls can be produced efficiently and could be responsible for baryon 
asymmetry \cite{SUSY} and dark matter \cite{SUSY-DM}. In the AD mechanism, there are two types of 
potentials: gravity-mediation type and gauge-mediation type. 
Here, we concentrate on the gravity-mediation type,
\beq\label{AD}
V_{AD}(\phi):=\frac{m^2}{2}\phi^2\left[
1+K\ln \left(\frac{\phi}{M}\right)^2
\right]~~~
{\rm with} ~~~ m^2,~M>0\ .
\eeq
In general, there may be nonrenormalizable terms, U(1) violation terms, and so on.
Here we neglect them for simplicity. 
Because Q-balls are typically supposed to be microscopic objects, their self-gravity is usually ignored.
Therefore, stability of Q-balls with various potentials has been intensively studied in flat spacetime 
\cite{stability,PCS01,SakaiSasaki,Copeland}. 
As for the AD potential (\ref{AD}), it has been known that equilibrium solutions for $K\geq 0$ are nonexistent while those  for $K<0$ are existent and stable. One may speculate that these properties are not changed by gravity if self-gravity is weak enough.

However, this speculation is not necessarily true for the following reasons.
First, for the potential $V=m^2\phi^2/2$, no equilibrium solution exists without gravity but equilibrium solutions, called (mini-)boson stars, exist due to self-gravity~\cite{boson-review}. 
This is a direct evidence that there are equilibrium solutions for $K=0$ with (\ref{AD}).

Second, in our previous paper~\cite{TamakiSakai2}, we considered gravitating Q-balls with 
\beq\label{V4}
V_4(\phi):={m^2\over2}\phi^2-\lambda\phi^4+\frac{\phi^6}{M^2} 
~~~{\rm with} ~~~ m^2,~\lambda,~M>0.
\eeq
In flat spacetime Q-balls with $V_4$ in the thick-wall limit are 
unstable and there is a minimum charge $Q_{{\rm min}}$, 
where Q-balls with $Q<Q_{{\rm min}}$ are nonexistent.
If we take self-gravity into account, on the other hand, there exist 
stable Q-balls with arbitrarily small charge, no matter how weak gravity is. 

Therefore, it is valuable to examine the influence of gravity in AD potential (\ref{AD}). 
As a result, we find that upper bound of the Q-ball charge appears due to gravity for $K<0$ 
and there appear ``Q-balls" for $K\geq 0$ which do not exist without gravity. 
Here we call all equilibrium solutions ``Q-balls" collectively, though solutions supported by gravity are usually called boson stars.

This paper is organized as follows.
In Sec. II, we derive equilibrium field equations. 
In Sec. III, we show numerical results of equilibrium Q-balls for $K<0$ 
and discuss existence of ``Q-balls" for $K\geq 0$. 
For $K=0$, we explain why ``Q-balls" called (mini-)boson stars exist with the influence of gravity.  
In the same mechanism, there appear ``Q-balls" even for $K>0$. In this case, ``Q-balls" 
are surrounded by Q-matter. 
In Sec. IV, we devote to concluding remarks.

\section{Analysis method of equilibrium Q-balls}

\subsection{Equilibrium field equations}

We begin with the action
\beq\label{Sg}
{\cal S}=\int d^4x\sqrt{-g}\left\{ \frac{{\cal R}}{16\pi G}-\frac12g^{\mu\nu}\pa_{\mu}\bp\cdot\pa_{\nu}\bp
-V(\phi)\right\}, 
\eeq
where $\bp=(\phi_1,~\phi_2)$ is an SO(2)-symmetric scalar field and 
$\phi:=\sqrt{\bp\cdot\bp}=\sqrt{\phi_1^2+\phi_2^2}$.
We assume a spherically symmetric and static spacetime, 
\beq\label{metric1}
ds^2=-\alpha^2(r)dt^2+A^2(r)dr^2+r^2(d\theta^2+\sin^2\theta d\varphi^2).
\eeq

For the scalar field, we assume that it has a spherically symmetric and stationary form, 
\beq\label{phase}
(\phi_1,\phi_2)=\phi(r)(\cos\omega t,\sin\omega t).
\eeq
Then the field equations become
\bea\label{Gtt}
-{r A^3\over2}G^t_t&:=&A'+{A\over2r}(A^2-1) \nonumber \\
&=&{4\pi G}r A^3\left({{\phi'}^2\over2A^2}
+{\omega^2\phi^2\over2\alpha^2}+V\right),
\\\label{Grr}
{r\alpha\over2}G_{rr}&:=&\alpha'+{\alpha\over2r}(1-A^2) \nonumber \\
&=&{4\pi G}r\alpha A^2
\left({{\phi'}^2\over2A^2}+{\omega^2\phi^2\over2\alpha^2}-V\right),
\\\label{Box}
{A^2\phi\over\phi_1}\Box\phi_1&:=&
\phi''+\left(\frac2r+{\alpha'\over\alpha}-{A'\over A}\right)\phi'
+\left({\omega A\over\alpha}\right)^2\phi \nonumber \\
&=&A^2{dV\over d\phi},
\eea
where $':= d/dr$. 
To obtain Q-ball solutions in curved spacetime, we should solve 
(\ref{Gtt})-(\ref{Box}) with boundary conditions, 
\bea
&& A(0)=A(\infty)=\alpha(\infty)=1,\nonumber \\ 
&& A'(0)=\alpha'(0)=\phi'(0)=\phi(\infty)=0.
\label{bcg}
\eea
We also restrict our solutions to monotonically decreasing $\phi (r)$. 
Because of the symmetry, there is a conserved charge called Q-ball charge,
\bea\label{Q}
Q&:= &\int d^3x\sqrt{-g}g^{0\nu}(\phi_1\pa_\nu\phi_2-\phi_2\pa_\nu\phi_1)=\omega I,
\nonumber  \\
&&{\rm where}~~~
I:=4\pi\int{A r^2\phi^2\over\alpha}dr.
\eea

We suppose $V_{AD}$ Model (\ref{AD}).
Rescaling the quantities as
\bea
&&\tp\equiv\frac{\phi}{M},~~
\tv_{AD}\equiv\frac{V_{AD}}{m^2M^2}=
\frac{\tp^2}{2}\left( 1+2K\ln \tp 
\right),~~ \nonumber  \\
&&\to\equiv\frac{\omega}{m},~~\kappa=GM^2,~~\tt\equiv mt,~~ \tr\equiv mr, 
\label{rescale}
\eea
the field equations (\ref{Gtt})-(\ref{Box}) are rewritten as
\beq\label{rsfe1}
A'+{A\over2\tr}(A^2-1)
=4\pi\kappa\tr A^3\left({\tpps\over2A^2}+{\to^2\tp^2\over2\alpha^2}+\tv_{AD}\right),
\eeq\beq\label{rsfe2}
\alpha'+{\alpha\over2\tr}(1-A^2)
=4\pi\kappa\tr\alpha A^2\left({\tpps\over2A^2}+{\to^2\tp^2\over2\alpha^2}-\tv_{AD}\right),
\eeq\beq\label{rsfe3}
\tp^{\prime\prime}+\left(\frac2{\tr}+{\alpha'\over\alpha}-{A'\over A}\right)\tpp
+\left({\to A\over\alpha}\right)^2\tp=A^2{d\tv_{AD}\over d\tp}.
\eeq

\subsection{Equilibrium solutions in flat spacetime}
\begin{figure}[htbp]
\psfig{file=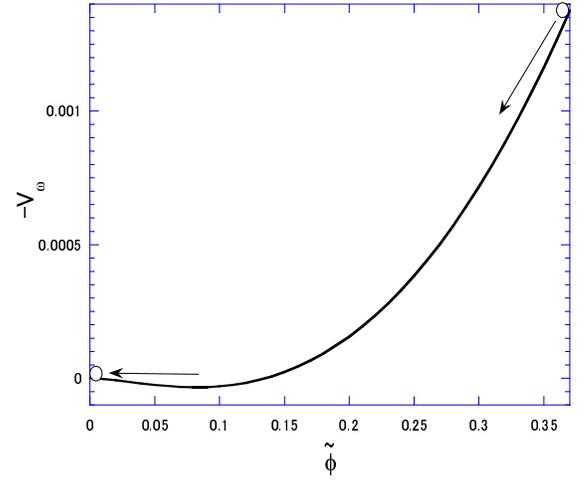,width=3in} 
\caption{$-V_{\omega}$ for a Q-ball in flat spacetime ($\kappa =0$). We put $K=-0.01$ and $\to^2 =1.04$. 
\label{flatV} }
\end{figure}

In preparation for discussing gravitating ``Q-balls", we review their 
equilibrium solutions in flat spacetime ($\kappa=0$).
The scalar field equation (\ref{rsfe3}) reduces to
\beq\label{rsfeflat}
\tp^{\prime\prime}=-\frac{2}{\tilde{r}}\tpp-\tilde{\omega}^2\tilde{\phi}+{d\tilde{V}_{AD}\over d\tilde{\phi}}\,.
\eeq
This is equivalent to the field equation for a single static scalar 
field with the potential $V_{\omega}:=\tilde{V}_{AD}-\tilde{\omega}^2\tilde{\phi}^2/2$.
Equilibrium solutions satisfying boundary conditions (\ref{bcg}) 
exist if 
\beq\label{excon}
{\rm max}(V_{\omega})>\tilde{V}_{AD}(\tp \ra 0)~~{\rm and}~~
{d^2V_{\omega}\over d\tilde{\phi}^2}(\tp \ra 0)>0.
\eeq
If we introduce $\epsilon^2 :=1-\tilde{\omega}^2$, we obtain 
\bea
\frac{dV_{\omega} }{d\tp}=\tp (\epsilon^2 +K+2K\ln \tp )\ , \label{origin1} \\
\frac{d^2 V_{\omega} }{d\tp^2}=\epsilon^2 +3K+2K\ln \tp \ . \label{origin2}
\eea
The second condition in (\ref{excon}) leads to
\beq\label{condition1}
K<0\ , 
\eeq
or 
\beq\label{condition1-2}
K=0\ {\rm and}\ \epsilon^2 >0\ . 
\eeq

In the former case (\ref{condition1}), $V_{\omega}$ has a maximum at 
$\tp =\tp_{1}:=e^{-\frac{\epsilon^2 +K}{2K}}$; then
the first condition in (\ref{excon}) becomes
\bea
\frac{K}{2}\tp_{1}^2 <0\ , \label{condition2}
\eea
which is trivially satisfied. 
In the latter case (\ref{condition1-2}), there is no maximum; then, there is no equilibrium solution.

If one regards the radius $r$ as \lq time\rq\ and the scalar amplitude $\phi(r)$ as \lq the position of a particle\rq, one can understand Q-ball solutions in words of Newtonian mechanics, as shown in Fig.\ 1.
Equation (\ref{rsfeflat}) describes a one-dimensional motion of a particle under the conserved force due to the potential $-V_{\omega}(\phi)$ and the \lq time\rq-dependent friction $-(2/r)d\phi/dr$.
Here we put $K=-0.01$, $\to^2 =1.04$. In this case, 
the scalar field $\tp \simeq 0.37$ at the initial time $\tr =0$ 
rolls down the potential and finally reaches $\tp =0$ at the time $\tr \ra \infty$.

\section{Gravitating ``Q-balls" }

The potential picture described above is also effective to argue equilibrium solutions in curved spacetime.
In this case,  $\epsilon^2$ should be redefined by 
\bea\label{thick2}
&&\epsilon^{2} :=1-\frac{\to^2}{\alpha^2} . \label{origin4} 
\eea
Because \lq the potential of a particle\rq, $-V_{\omega}$, is now  \lq time\rq-dependent, 
the existence conditions of equilibrium solutions are not as simple as those in flat spacetime.

\subsection{$K<0$}
\begin{figure}[htbp]
\psfig{file=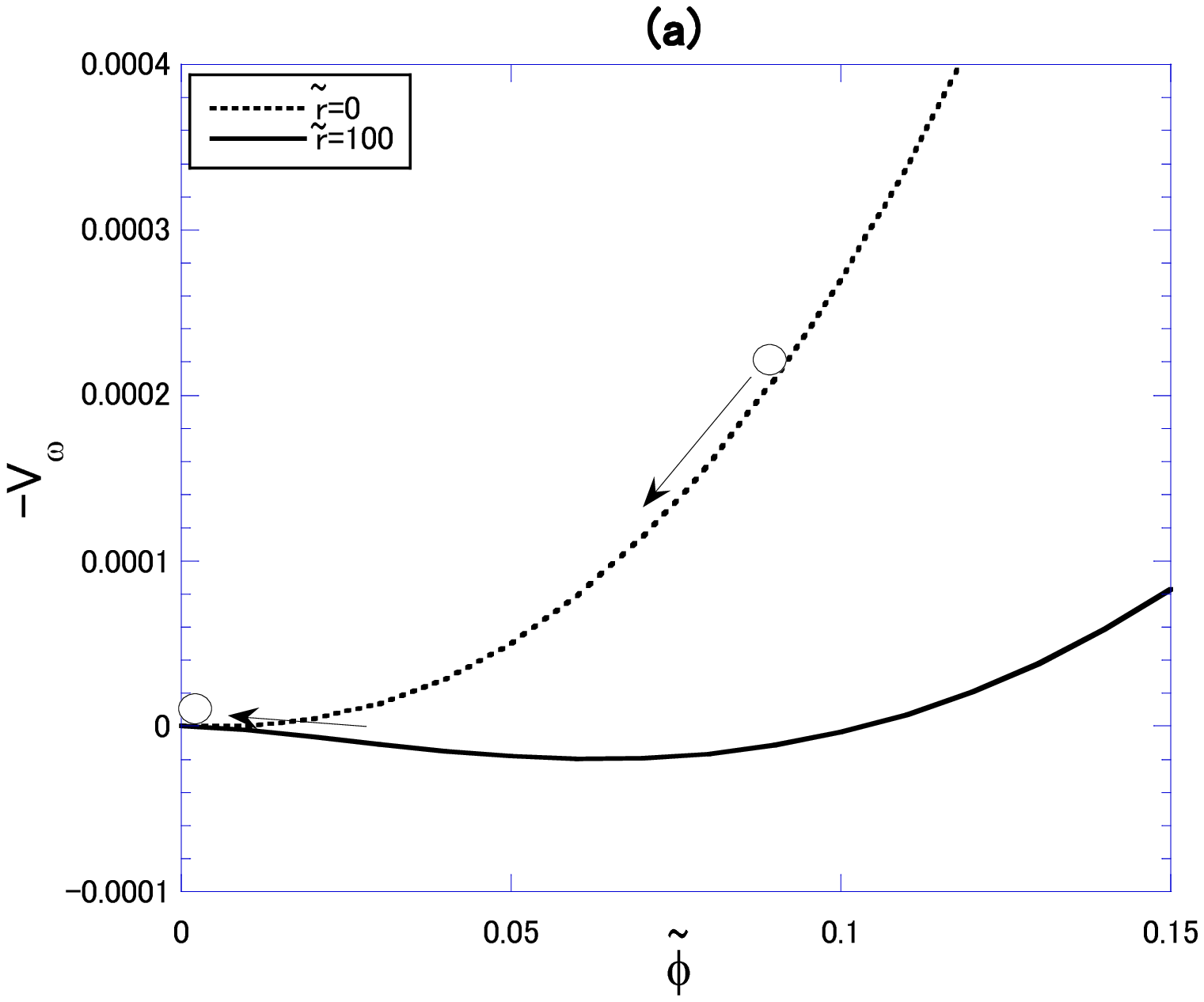,width=3in} \\
\psfig{file=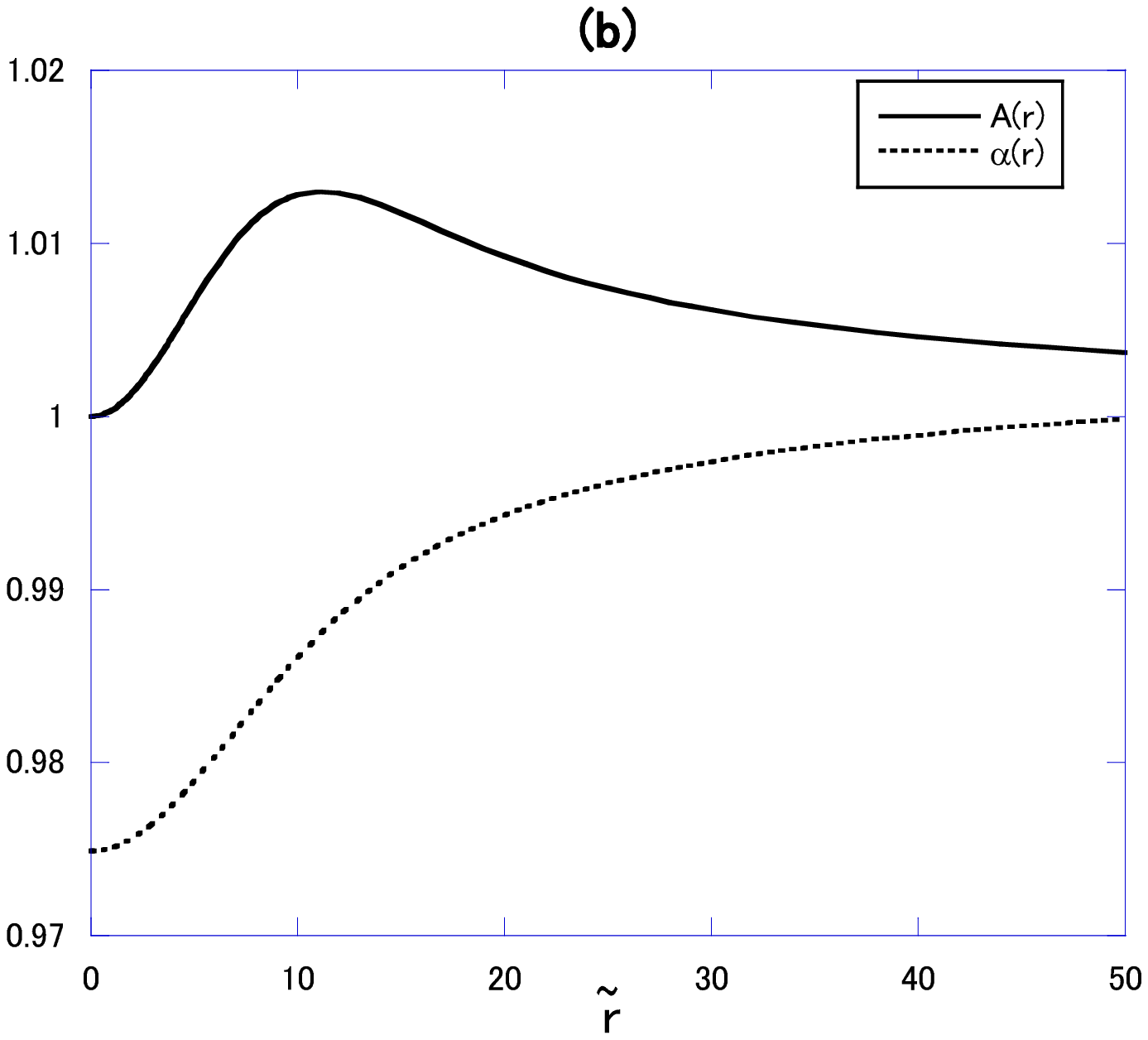,width=3in}
\caption{(a) $-V_{\omega}$ and (b) behaviors of the metric functions 
for a gravitating Q-balls. We put $K=-0.01$, $\to^2 \simeq 1.045$ and $\kappa =0.01$. 
$V_{\omega}$ changes as \lq time\rq\ $\tr$ goes. 
\label{VK-001} }
\end{figure}
\begin{figure}[htbp]
\psfig{file=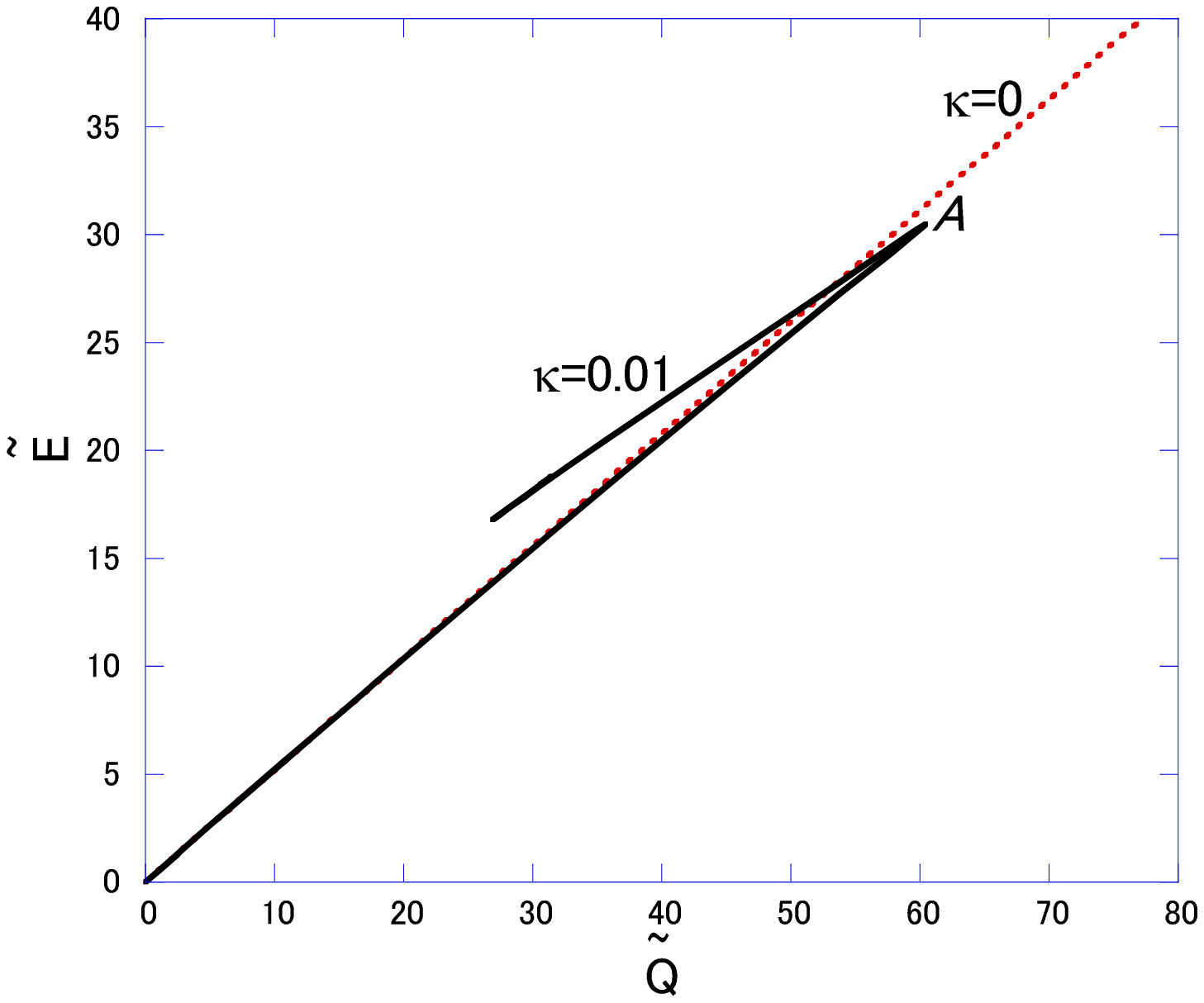,width=3in} 
\caption{$\tQ$-$\tE$ relation for $K=-0.01$. 
\label{QE} }
\end{figure}
We discuss the existence of equilibrium solutions by analogy with Newtonian mechanics, 
as shown in Fig.\ \ref{VK-001} (a). We also exhibit behaviors of 
the metric functions in Fig.\ \ref{VK-001} (b). 
Because $-V_{\omega}$ depends on the \lq time\rq 
$\tr$, it has a minimum at $\tp \simeq 0.08$ when $\tr =100$ while it has a minimum at $\tp\simeq0$ 
when $\tr =0$. 
At the \lq initial time\rq\ $\tr =0$ the scalar field at $\tp\simeq 0.089$ 
rolls down the potential and finally reaches $\tp =0$ at the time $\tr \ra \infty$.
We thus understand how gravity changes properties of equilibrium solutions.

As we discussed in our previous papers~\cite{TamakiSakai2,TamakiSakai}, stability of 
Q-balls can be easily understood from the relation between $Q$ and 
the Hamiltonian energy $E$, which is defined by 
\beq\label{H}
E=\lim_{r\ra\infty}{r^2\alpha'\over2GA}={M_S\over2},
\eeq
where $M_S$ is the Schwarzschild mass. 
Here, stability means local stability, that is, stability against small perturbations. 
We also normalize $E$ and $Q$ as
\beq\label{Enormalize}
\tE := \frac{mE}{M^2},~~ \tQ :=\frac{m^2 Q}{M^2}.
\eeq

We compare $\tQ$-$\tE$ relation for the flat case ($\kappa =0$) 
with that for the gravitating case $\kappa =0.01$ in Fig.~\ref{QE}. 
In the case that $\kappa=0$, $\tQ$ is almost proportional to $\tE$, and accordingly, 
all solutions for this parameter range are stable.
In the case that $\kappa=0.01$, however, there is a cusp structure at the point $A$, 
where stability changes. 
If there are two solutions for fixed $\tQ$, the solution with larger energy $\tE$ should be  unstable.
That is, the upper branch represents unstable solutions.
At the same time, this cusp structure indicates that there is a maximum charge $\tQ_{{\rm max}}$, where 
solutions with $\tQ>\tQ_{{\rm max}}$ are nonexistent due to gravity.
This is a common feature with $V_{3}$ and $V_{4}$ models~\cite{TamakiSakai2,TamakiSakai,multamaki}. If we take larger (smaller) $\kappa$, 
the point corresponding to $A$ has smaller (larger) $\tE$. 
However, qualitative features do not change. 

\subsection{$K=0$}
\begin{figure}[htbp]
\psfig{file=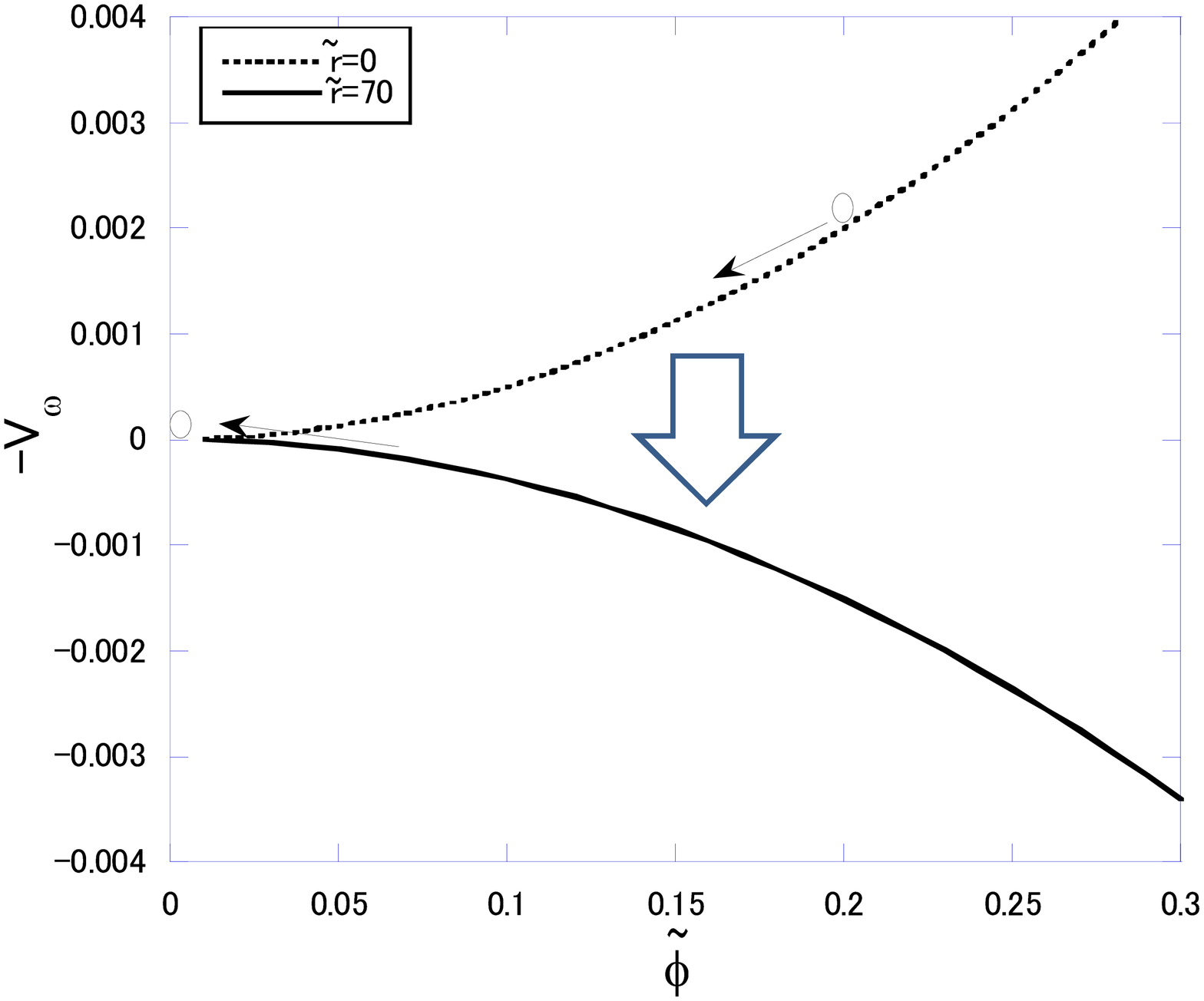,width=3in} 
\caption{$-V_{\omega}$ for a (mini-)boson star. We put $K=0$, $\to^2 \simeq 0.92$ and $\kappa =0.01$. 
Because $-V_{\omega}$ changes as $\tr$, the scalar field with $\tp \simeq 0.2$ at the initial time $\tr =0$ 
rolls down and finally reaches $\tp =0$ at the time $\tr \ra \infty$.
\label{VK0} }
\end{figure}
\begin{figure}[htbp]
\psfig{file=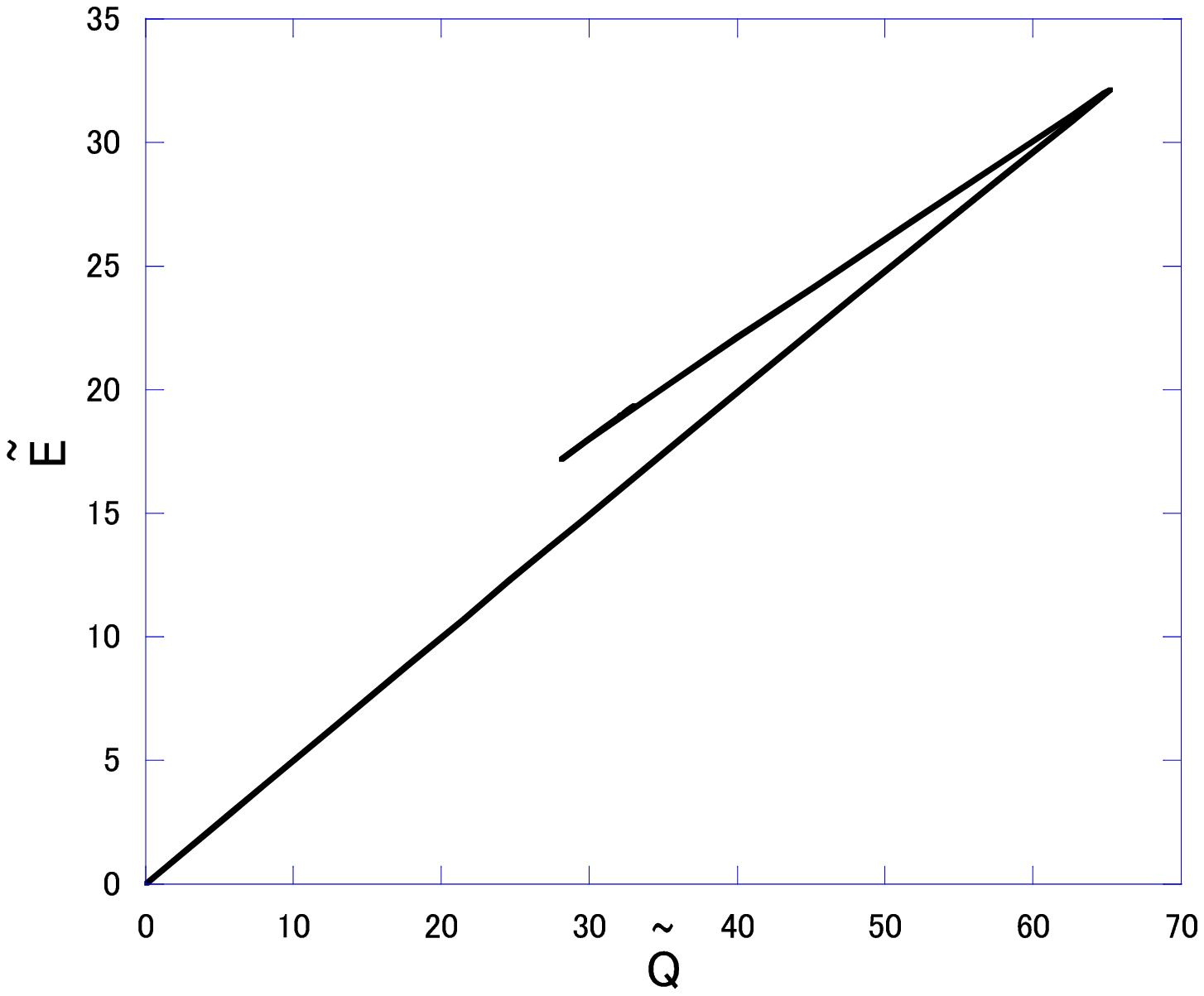,width=3in} 
\caption{$\tQ$-$\tE$ relation for $K=0$ and $\kappa =0.01$. 
\label{QEK0} }
\end{figure}

This case corresponds to the potential for the mini-boson stars, 
which have been investigated in the literatures~\cite{boson-review}. 
First, we explain why mini-boson stars appear if we include self-gravity. 
The key point is that  the sign of $\epsilon^2=d^2\tv/d\tp^2(0)$ depends on $\tr$.
Figure \ref{VK0} shows how the shape of $-V_{\omega}$ changes as $\tr$ varies.
The scalar field rolls down the potential $-V_{\omega}$ near $\tr =0$ 
while it climbs up $-V_{\omega}$ in the asymptotic region. 
As a result, the scalar field at the \lq initial position\rq\ $\tp (0)\simeq 0.2$ satisfies the asymptotic boundary
condition $\tp (\infty )=0$. 
In contrast, in the case of flat spacetime, because $-V_{\omega}$ is a monotonically decreasing (increasing) function of $\tp$ for $\epsilon^2 >0$ ($<0$), there is no equilibrium  solution. 

We also show $\tQ$-$\tE$ relation for $K=0$ and $\kappa =0.01$ in Fig.~\ref{QEK0}. 
The result is similar to that for $K=-0.01$. 
There is a maximum charge $\tQ_{{\rm max}}$.
The lower branch represents stable solutions while the upper branch unstable.

\subsection{$K>0$}
\begin{figure}[htbp]
\psfig{file=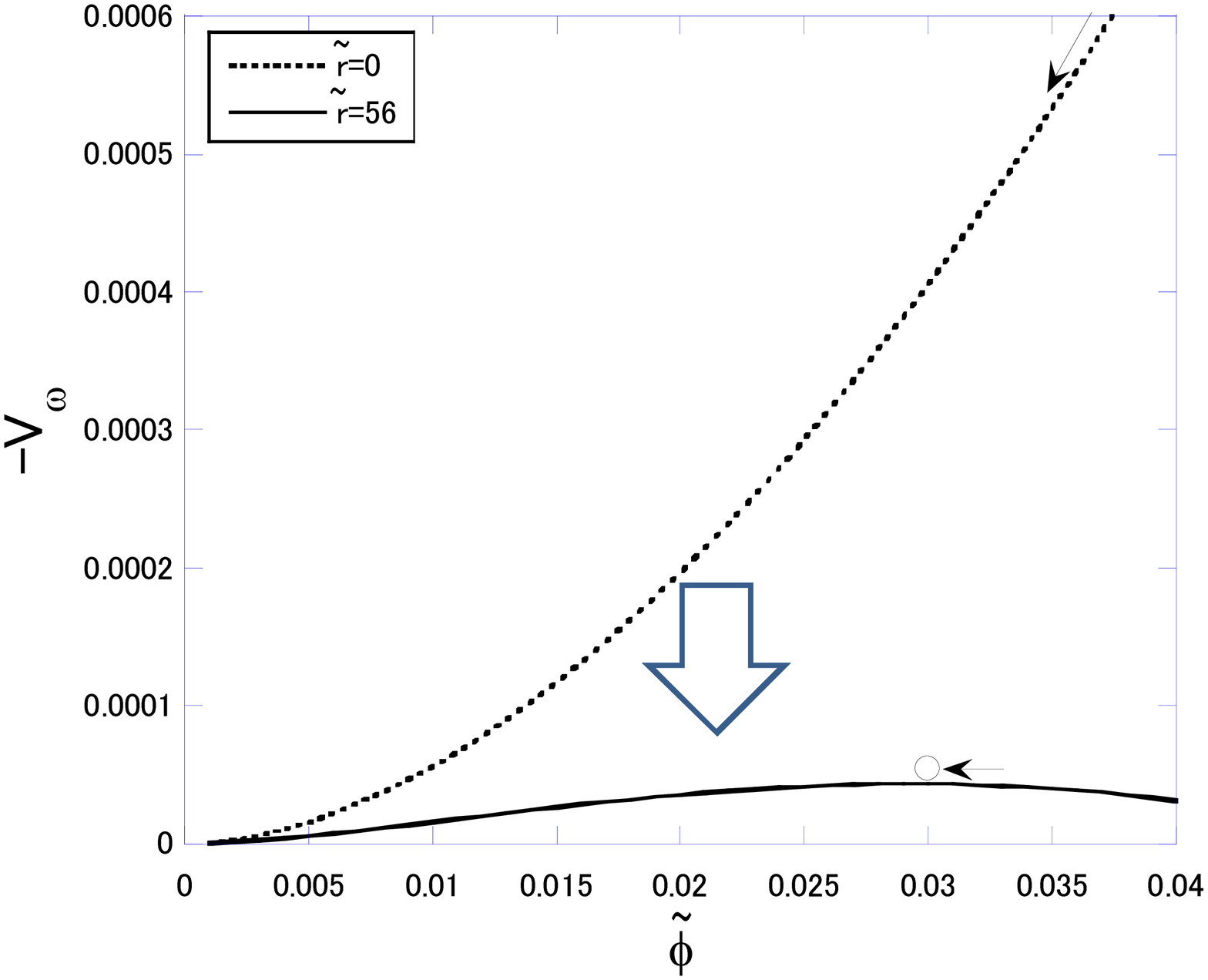,width=3in} 
\caption{$-V_{\omega}$ for $K>0$.
We put $K=0.1$, ${\to^2}/{\alpha^2}(\tr=0)\simeq 1.2$, and $\kappa =0.01$.
Because of $d^2V_{\omega}/d\tilde{\phi}^2(\tp \ra 0)<0$, 
there is no Q-ball which satisfies (\ref{bcg}). However, the scalar field can stop at 
the maximum of $-V_{\omega}$, $\tp =\tp_{1}\simeq 0.03$ in the large $\tr$ region, 
if the scalar field with initial value $\tp (0)\simeq 0.56$  (outside the figure) rolls down.
\label{VK01} }
\end{figure}
\begin{figure}[htbp]
\psfig{file=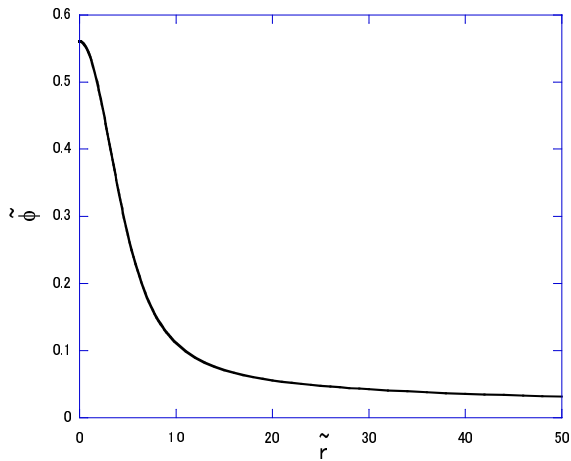,width=3in} 
\caption{Configuration of the scalar field for $K=0.1$, 
${\to^2}/{\alpha^2}(\tr=0)\simeq 1.2$ around $\tr =0$ and $\kappa =0.01$. 
\label{config} }
\end{figure}
As in the case that $K=0$, the shape of $V_{\omega}$ depends on $\tilde{r}$.
Figure~\ref{VK01} shows the potential $-V_{\omega}$ for $K=0.1$, $\frac{\to^2}{\alpha^2} \simeq 1.2$ 
at $\tr =0$ and $\kappa =0.01$. 
However, because $d^2V_{\omega}/d\tilde{\phi}^2(\tp \ra 0)<0$ regardless of $\epsilon^2$, there is no 
Q-ball solution which satisfies (\ref{bcg}) even if we include gravity. 

However, we should notice $-V_{\omega}$ maximum around $\tp =\tp_{1}\simeq 0.03$ in 
the large $\tr$ region. Figure \ref{VK01} indicates that if the scalar field rolls down 
from $\tp >\tp_{1}$, there is a solution which satisfies $\tp =\tp_{1}$ at $\tr\ra\infty$. 
We show the example of such a solution in Fig.~\ref{config} for the same parameters as in 
Fig.~\ref{VK01}. We have also confirmed that this kind of solution is generic for $K>0$.

Our solutions are not asymptotically flat but surrounded by Q-matter.
Because the energy $E$ and the charge $Q$ are diverging, we cannot apply energetics or catastrophe theory to these solutions.
Stability analysis is the next important issue.

\section{Conclusion and discussion}

We have investigated gravitating ``Q-balls" in the gravity-mediated AD mechanism (\ref{AD}).
Contrary to the flat case, in which equilibrium solutions exist only if $K<0$, we have found three types of gravitating solutions as follows.
In the case that $K<0$, ordinary Q-ball solutions exist; there is an upper bound of the charge due to gravity.
In the case that $K=0$, equilibrium solutions called (mini-)boson stars appear due to gravity; there is an upper bound of the charge, too.
In the case that $K>0$, equilibrium solutions appear, too. In this case, 
these solutions are not asymptotically flat but surrounded by Q-matter.
It is worth noting that because the amplitude of the scalar field can grow even for $K>0$~\cite{Mazumdar}, 
our solution may play an important role as dark matter. 

Our present work as well as previous work \cite{TamakiSakai2,TamakiSakai} suggests that self-gravity 
may change properties of the solutions even if it is weak.
Therefore, it may be important to extend our approach to other models such as the gauge-mediation type.

\acknowledgements
We would like to thank Kei-ichi Maeda for continuous encouragement. 
The numerical calculations were carried out on SX8 at  YITP in Kyoto University. 
This work was supported by MEXT Grant-in-Aid for Scientific Research on Innovative Areas No.\ 22111502.


\end{document}